\documentclass[sigconf, 9pt]{acmart}
\AtBeginDocument{%
  }
\setcopyright{none}
\copyrightyear{}
\acmYear{}
\acmDOI{}
\acmISBN{}
\acmConference[ISLPED '26]{IEEE/ACM International Symposium on Low Power Electronics and Design}{August 05--07,
  2026}{Evanston, IL}

\usepackage{amsmath,amsfonts}
\usepackage{algorithm}
\usepackage{algorithmic}
\usepackage{graphicx}
\usepackage{textcomp}
\usepackage{xcolor}
\usepackage{subcaption}

\usepackage{colortbl}
\usepackage{dhucs} 
\usepackage{booktabs} 
\usepackage{multirow}
\usepackage{comment}
\usepackage{lineno}
\usepackage{color} 
\usepackage{setspace} 
\usepackage{makecell}
\usepackage{enumitem} 
\usepackage{relsize}
\usepackage{tabularx}
\usepackage{diagbox}
\usepackage{bm}
\usepackage{hhline}
\usepackage{tikz}
\usepackage{pgfplots}
\usepackage{xurl}
\usepackage{needspace} 
\usepackage{stfloats}

\newcommand\LKW[1]{\textcolor{black}{#1}}
\newcommand\JJM[1]{\textcolor{black}{#1}}

\newcommand\korean[1]{}

\newcommand\refFigure[1]{Figure~\ref{#1}}

\newcommand\refEqn[1]{(\ref{#1})}

\newcommand\blind[1]{XXXX}

\makeatletter
\newcommand{\myspacedsection}[1]{%
  \@startsection{section}{1}{\z@}%
    {6pt}
    {2pt}
    {\normalfont\Large\bfseries}%
  {#1}}
\makeatother

\renewcommand{\baselinestretch}{1}
\begin{document}

\title{SA-Kura: An Energy-Efficient Systolic Array Accelerator for Locally-Coupled Kuramoto Drift in Diffusion Sampling}

\author{Jeongmin Jin, Kyeongwon Lee, Mundo Jeong, Jongin Choi, and Woojoo Lee}
\affiliation{%
  \institution{
}
  \city{}
  \country{} 
}
\email{}

\thanks{This paper has been accepted for publication in the Proceedings of the ACM/IEEE International Symposium on Low Power Electronics and Design (ISLPED), 2026.}

\thanks{
This work was supported in part by the National Research Foundation of Korea (NRF) grant funded by MSIT (No. RS-2024-00345668), and in part by Institute of Information \& communications Technology Planning \& Evaluation (IITP) grants funded by MSIT (No. RS-2023-00277060).

Jeongmin Jin and Kyeongwon Lee contributed equally to this work. 

Woojoo Lee is the corresponding author.}

\begin{abstract}
Diffusion inference remains costly for edge deployment, yet existing accelerators focus almost exclusively on score networks because standard drift is merely a trivial linear scaling. 
Kuramoto orientation diffusion replaces this trivial drift with locally coupled phase interactions, improving sampling efficiency but introducing a new hardware bottleneck: a center-dependent nonlinear $5\times5$ stencil evaluated at every reverse step. 
This kernel maps poorly to conventional CNN accelerators and matrix-oriented engines. 
We present SA-Kura, to our knowledge the first digital systolic-array accelerator dedicated to locally coupled Kuramoto drift. 
By reformulating pairwise sinusoidal coupling into neighbor accumulation independent of the center phase followed by a single center-dependent multiply--subtract combination, SA-Kura eliminates in-PE transcendental units and enables regular systolic execution with register-level reuse. 
SA-Kura was implemented in synthesizable RTL, integrated into a lightweight RISC-V-based SoC, prototyped on FPGA, and evaluated through 45\,nm CMOS synthesis and power analysis. 
For the drift kernel only, compared with software execution of the same kernel on the processor core in the same SoC platform, SA-Kura reduces latency and energy by $193\times$ and $69.4\times$, respectively. Compared with a standalone Jetson Orin Nano CUDA implementation of the same kernel, it is $6.57\times$ faster and achieves approximately $46.0\times$ lower energy per pixel.

\end{abstract}

\maketitle

\section{Introduction}

Diffusion models have become a major workload in generative AI, but their iterative reverse-time inference remains expensive for energy-constrained edge deployment~\cite{Guo:ISSCC24,Kong:ISCA25,Ding:ASPDAC25,Park:DAC25,Kim:JSSC26}. 
Starting from near-pure noise, a sample is generated by repeatedly evaluating a score network and updating the current state. 
Because this procedure is inherently sequential, both latency and energy grow almost linearly with the number of sampling steps~\cite{Karras:NIPS22,Salimans:ICLR22}. 
Early diffusion formulations therefore often require hundreds to thousands of reverse steps for high-quality generation~\cite{Ho:NeurIPS20,Song:ICLR21,Luo:ICLR24}. 
To reduce this cost, prior work has largely pursued two directions: lowering the cost of each step through score-network compression, feature reuse, and dedicated accelerators~\cite{Kim:ECCV24,Li:ICCV23,Ma:CVPR24,Wimbauer:CVPR24,Guo:ISSCC24,Kong:ISCA25,Ding:ASPDAC25,Park:DAC25,Kim:JSSC26}, and lowering the number of steps through improved samplers and distillation~\cite{SongJ:ICLR21,Lu:NeurIPS22,Song:ICML23,Ma:TPAMI25}. 
Most of this literature, however, assumes the standard variance-preserving stochastic differential equation (VP-SDE), whose drift term is a trivial linear scaling and is therefore usually treated as computationally negligible~\cite{Song:ICLR21}.

Recently, Kuramoto orientation diffusion challenged this assumption by replacing the trivial drift with locally-coupled sinusoidal phase interactions~\cite{Song:NeurIPS25}. 
This change is algorithmically important because it preserves local structure longer during the forward process and improves step efficiency on orientation-rich data. 
In particular, the original work reported that a 100-step Kuramoto model outperformed a 1000-step score-based generative model on the Brodatz texture dataset~\cite{Song:NeurIPS25}. 
From a hardware perspective, this result is equally significant: once drift becomes a nontrivial local operator rather than a negligible linear term, it emerges as a repeated kernel that must be accelerated in its own right.

The locally-coupled Kuramoto drift computes, at every pixel and every sampling step, a $5\times5$ neighborhood reduction over center-dependent phase differences of the form $\sin(\theta_j-\theta_i)$. 
Although its arithmetic count is much smaller than that of a U-Net, the kernel is fundamentally mismatched to conventional matrix-oriented engines. 
Because the operation depends on the center phase $\theta_i$, it cannot be reduced to a fixed-weight convolution, an im2col transform, or a single GEMM. 
As a result, GPU tensor cores and MAC-centric CNN accelerators cannot exploit their usual strengths on this kernel. 
This is the key hardware observation of this work: even a kernel with modest arithmetic count can become a practical bottleneck when its computation pattern is poorly matched to the underlying datapath. 
Our kernel-level profiling on Jetson Orin Nano shows that the drift kernel alone accounts for approximately 16.5\% of the latency of a sampling step. 
This residual cost becomes even more important as diffusion accelerators continue to reduce score-network latency~\cite{Guo:ISSCC24,Kong:ISCA25}. 
Without dedicated drift support, the end-to-end pipeline can still stall on drift and lose a significant fraction of the score-side acceleration benefit.


Existing hardware for phase-coupled dynamics does not directly address this setting. 
Analog and mixed-signal oscillator systems are difficult to scale to image-sized lattices because they map one physical oscillator to one computational node~\cite{Moy:NatureElec22,Delacour:NCE23,Todri:npjUC24}. 
Prior digital oscillatory neural network (ONN) accelerators target relatively small fully connected graphs or specialized optimization problems rather than streaming image-scale stochastic updates~\cite{Bashar:ISQED24,Haverkort:Neuroscience25}. 
Conventional convolution accelerators are also inadequate because they assume input-independent kernel weights~\cite{Chen:ISCA16,Shafiee:ISCA16}, whereas Kuramoto drift requires center-dependent nonlinear post-processing after neighborhood access. 
What is missing, therefore, is a scalable digital accelerator that can exploit overlapping-neighborhood reuse while efficiently supporting image-scale local phase coupling under tight energy and area constraints.

This paper presents SA-Kura, to the best of our knowledge, the first digital systolic-array accelerator for locally-coupled Kuramoto drift in diffusion sampling. 
SA-Kura is designed as a dedicated drift coprocessor that operates alongside a host processor or a separate score-network accelerator. 
Because both score and drift are computed from the same current state, this organization allows drift computation to proceed in parallel with score evaluation and prevents drift from remaining a serialized residual bottleneck. 
At the algorithm-to-hardware level, SA-Kura reformulates the pairwise sinusoidal coupling using a trigonometric identity, converting the original center-dependent stencil into neighbor accumulation independent of the center phase, followed by a single center-dependent multiply--subtract combination. 
This removes transcendental units from individual processing elements (PEs) and exposes a regular dataflow suitable for systolic execution. 
SA-Kura further integrates a quarter-wave LUT with linear interpolation for single-cycle $\sin/\cos$ generation, a two-dimensional offset-sweep dataflow for register-level reuse of overlapping neighborhoods, and drain--prefill overlap to eliminate tile-boundary idle cycles. 
The PE array is parameterized by $(N_h,N_w)$, where $N_h$ and $N_w$ denote the numbers of PE rows and columns, respectively; the array shape jointly determines throughput, utilization, and energy efficiency.

We implement SA-Kura in synthesizable RTL, integrate it into a lightweight SoC platform, and evaluate it across 25 array configurations in a 45\,nm CMOS flow. 
Among the synthesized configurations, an asymmetric $20\times5$ array emerges as the system-level optimum, offering both high throughput and the lowest measured system-level energy at 5.88\,nJ/px.
Compared with software execution of the same fixed-point drift kernel on the processor core within the same SoC platform, the $20\times5$ configuration reduces latency and energy by $192.99\times$ and $69.39\times$, respectively. 
Relative to a Jetson Orin Nano CUDA baseline, it is approximately $6.57\times$ faster and achieves approximately $46.02\times$ lower energy per pixel.

The main contributions of this work are as follows:
\begin{itemize}[leftmargin=*, nosep, topsep=0pt]
\item We identify locally-coupled Kuramoto drift as a new hardware bottleneck in step-efficient diffusion models, and reformulate its center-dependent stencil into neighbor accumulation independent of the center phase followed by a single multiply--subtract combination, thereby eliminating in-PE transcendental units.
\item We propose SA-Kura, a parameterized digital systolic-array accelerator that combines shared LUT-based $\sin/\cos$ generation, a 2-D offset-sweep dataflow, in-dataflow center capture, and drain--prefill overlap for efficient neighbor reuse.
\item We implement SA-Kura in synthesizable RTL, integrate it into a lightweight SoC platform, validate it across 25 FPGA/ASIC configurations, derive accurate closed-form cycle and power/area models, and show substantial benefits over both a software baseline on the same platform and a Jetson Orin Nano baseline.
\end{itemize}


\section{Kuramoto Drift: Modeling and Hardware Reformulation}
\label{sec:preliminary}


\subsection{Kuramoto Drift in Score-Based Diffusion}
\label{sec:prelim_kuramoto}

Score-based diffusion models generate samples by defining a forward corruption process and learning its reverse-time dynamics~\cite{Song:ICLR21}. 
In continuous time, the forward and reverse processes are given by
\begin{align}
\text{Forward:}\quad & dx = f(x,t)\,dt + g(t)\,dw,
\label{eq:forward_sde} \\
\text{Reverse:}\quad & dx = \bigl[f(x,t) - g^2(t)\,\nabla_x \log p_t(x)\bigr]\,dt + g(t)\,d\bar{w},
\label{eq:reverse_sde}
\end{align}
where $f(x,t)$ is the drift term, $g(t)\,dw$ is the diffusion term, and $\nabla_x \log p_t(x)$ is the score function that guides the noisy state toward the data distribution. 
In practice, the score is approximated by a neural network, typically a U-Net~\cite{Song:ICLR21}. 
In the standard variance-preserving SDE (VP-SDE), the drift takes the form $f(x,t)=-\tfrac{1}{2}\beta(t)\,x$, which is a computationally trivial isotropic contraction. 
Because this drift acts independently on each state variable and attenuates local structure isotropically, the reverse process typically requires many sampling steps to recover high-fidelity samples.

Kuramoto orientation diffusion replaces this trivial drift with nonlinear phase-coupled dynamics~\cite{Song:NeurIPS25}. 
In this model, each pixel is interpreted as a phase variable in $[-\pi,\pi)$, and the locally-coupled variant considered in this work evolves the phase at location $i$ according to
\begin{equation}
d\theta_t^i =
\left[
\frac{K(t)}{|\mathcal{N}_i|}
\sum_{j \in \mathcal{N}_i} \sin(\theta_t^j - \theta_t^i)
+
K_{\text{ref}}(t)\sin(\psi_{\text{ref}} - \theta_t^i)
\right]dt
+
\sqrt{2D_t}\,dW_t^i,
\label{eq:lc_kuramoto}
\end{equation}
where $\mathcal{N}_i$ denotes the $5\times5$ neighborhood of pixel $i$, $K(t)$ is the local coupling strength, $K_{\text{ref}}(t)$ and $\psi_{\text{ref}}$ are the reference gain and phase, $D_t$ is the diffusion coefficient, and $W_t^i$ is an independent Wiener process. 
The first term in the bracket is the local neighborhood coupling, the second term attracts the phase field toward the global reference phase $\psi_{\text{ref}}$, and the stochastic term injects diffusion noise. 
Unlike the VP-SDE drift, this locally-coupled interaction preserves local orientation information through anisotropic phase alignment, producing a structured corruption process rather than an isotropic decay. 
This structured corruption underlies the improved step efficiency reported for orientation-rich data in prior work~\cite{Song:NeurIPS25}. 
From a hardware perspective, this observation is equally important: the drift is no longer negligible, but instead becomes a repeated local kernel and an acceleration target orthogonal to score-network computation.

\begin{figure}[t]
\centering
\includegraphics[width=\linewidth]{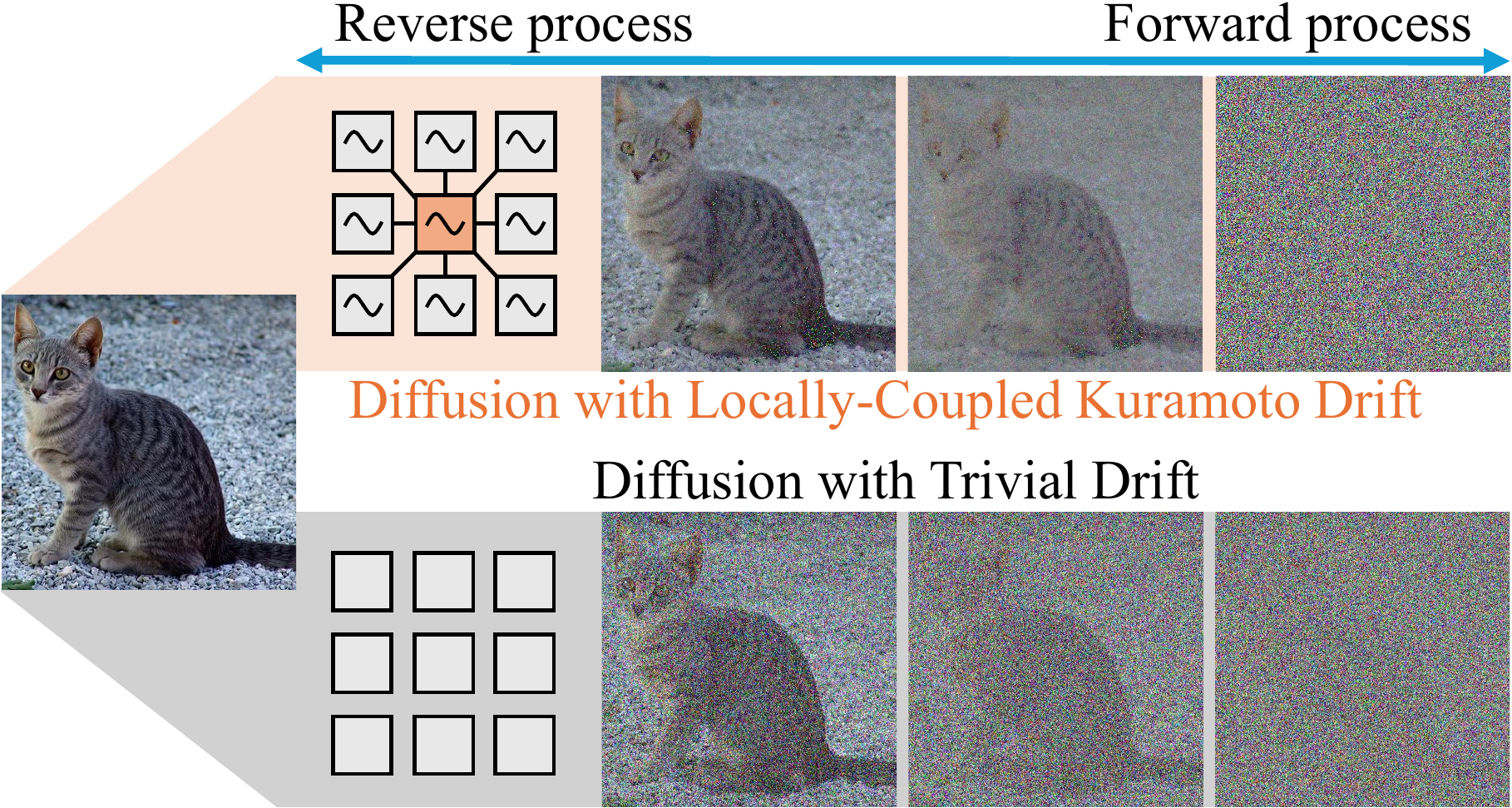}
\vskip -6pt
\caption{Diffusion trajectories under Kuramoto drift (top) and trivial drift (bottom). Kuramoto drift preserves stripe-like local structure longer during forward corruption.}\label{fig:kuramoto_diffusion_concept}
\vskip -2pt
\end{figure}

\refFigure{fig:kuramoto_diffusion_concept} qualitatively compares forward corruption under locally-coupled Kuramoto drift and trivial drift. 
Under locally-coupled drift, stripe-like structures remain visible for substantially longer than under trivial drift, leaving richer cues for the reverse process. 
This qualitative behavior is consistent with the improved sampling efficiency reported for Kuramoto diffusion and motivates dedicated hardware support for its repeated local phase-coupling computation. 
In SA-Kura, the dominant acceleration target is the neighborhood-coupling term, while the reference-attraction and stochastic terms are incorporated outside the PE array after neighborhood reduction.

\subsection{Hardware-Oriented Reformulation of Local Phase Coupling}
\label{sec:prelim_optimization}

For notational simplicity, we omit the explicit time index $t$ in the remainder of this subsection.

A direct hardware implementation of the neighborhood-coupling term in \eqref{eq:lc_kuramoto} is inefficient. 
For each center pixel $i$, the neighborhood term requires $|\mathcal{N}_i|=25$ evaluations of $\sin(\theta_j-\theta_i)$. 
For a $96\times96$ image, this corresponds to $96\times96\times25 = 230{,}400$ pairwise interactions, and a direct implementation therefore entails the same number of transcendental evaluations per sampling step. 
This creates three hardware difficulties. 
First, assigning a transcendental unit such as CORDIC or a dedicated LUT to every processing element (PE) is area- and power-intensive. 
Second, because $\sin(\theta_j-\theta_i)$ depends on the center phase $\theta_i$, the computation cannot be reduced to a standard fixed-weight convolution, an im2col transform, or a single GEMM, limiting the usefulness of GPU tensor cores and MAC-centric accelerator arrays. 
Third, although adjacent centers share most of their $5\times5$ neighborhoods, direct evaluation does not expose reusable partial results because the contribution of the same neighbor still changes with the center phase.

The key observation is that the pairwise coupling can be algebraically separated into neighbor accumulation independent of the center phase, followed by a single center-dependent multiply--subtract combination:
\begin{equation}
\sin(\theta_j-\theta_i)
=
\sin\theta_j\,\cos\theta_i
-
\cos\theta_j\,\sin\theta_i.
\label{eq:prelim_trig_identity}
\end{equation}
Define the neighborhood accumulations as
\begin{equation}
S_i \triangleq \sum_{j\in\mathcal{N}_i}\sin\theta_j,\qquad
C_i \triangleq \sum_{j\in\mathcal{N}_i}\cos\theta_j.
\label{eq:prelim_sc_simple}
\end{equation}
Then the unscaled neighborhood core becomes
\begin{equation}
\tilde{u}_i^{\mathrm{nbr}}(\theta)
\triangleq
\cos\theta_i \cdot S_i - \sin\theta_i \cdot C_i,
\label{eq:prelim_nbr_core}
\end{equation}
and the neighborhood-coupling term in \eqref{eq:lc_kuramoto} can be written as
\begin{equation}
u_i^{\mathrm{nbr}}(\theta,t)
=
\frac{K(t)}{|\mathcal{N}_i|}
\tilde{u}_i^{\mathrm{nbr}}(\theta).
\label{eq:prelim_center_combine}
\end{equation}
In this work, $\mathcal{N}_i$ denotes the full $5\times5$ window including the center pixel. 
In hardware, however, the center phase is captured separately for the final center-dependent combination and is not accumulated as a neighbor. 
This does not change \eqref{eq:prelim_center_combine}, because the omitted self contribution cancels exactly as
$\cos\theta_i\sin\theta_i-\sin\theta_i\cos\theta_i=0$.

The reference-attraction term can be expressed in the same form:
\begin{equation}
u_i^{\mathrm{ref}}(\theta,t)
=
K_{\text{ref}}(t)
\bigl(
\sin\psi_{\text{ref}}\cos\theta_i
-
\cos\psi_{\text{ref}}\sin\theta_i
\bigr).
\label{eq:prelim_ref_combine}
\end{equation}
Accordingly, both deterministic components of the drift reuse the same center quantities $\sin\theta_i$ and $\cos\theta_i$, while only the neighborhood term requires a spatial reduction over $\mathcal{N}_i$.

\begin{figure*}[t]
    \vskip -3pt
    \centering
    \includegraphics[width=0.98\textwidth]{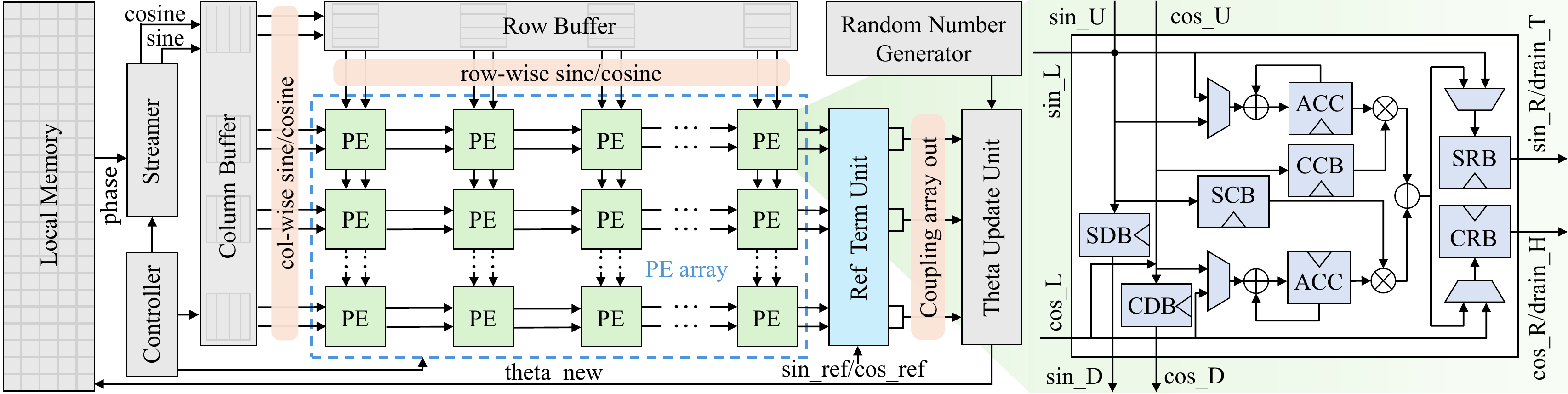}
    \vskip -6pt
\caption{Overview of SA-Kura, a parameterized systolic-array accelerator for locally-coupled Kuramoto drift. The left panel shows the system-level architecture, including the streaming, buffering, PE array, and drift-update path with reference and stochastic terms, while the right panel details the PE microarchitecture.}
    \label{fig:sa_kura_overview}
    \vskip -6pt
\end{figure*}

This reformulation does not remove transcendental evaluation altogether; rather, it relocates it to a more hardware-efficient point in the datapath. 
Instead of evaluating $\sin(\theta_j-\theta_i)$ for every pair $(i,j)$, each streamed phase value is first converted into $\sin\theta$ and $\cos\theta$ before entering the PE array, and the array itself performs only accumulation, multiplication, and subtraction. 
The PE computation is therefore decomposed into two regular stages:
1) a $5\times5$ offset sweep in which the center offset is captured and the remaining $|\mathcal{N}_i|-1$ neighbor offsets are accumulated into $S_i$ and $C_i$, and
2) a final center-dependent multiply--subtract combination using $\cos\theta_i$ and $\sin\theta_i$. 
This shifts transcendental evaluation out of the per-PE inner loop, removes the need for local transcendental units inside the PE array, and exposes a regular computation pattern that is well matched to systolic execution.


The reformulation also reveals the key reuse opportunity exploited by SA-Kura. 
Because $S_i$ and $C_i$ depend only on neighboring values, transformed inputs $\sin\theta_j$ and $\cos\theta_j$ can be reused across adjacent centers whose neighborhoods overlap spatially. 
In the direct phase-difference form, this reuse is obscured by the center dependence of $\sin(\theta_j-\theta_i)$; after reformulation, it becomes explicit and can be captured through register-level data movement.

Although the accumulation stage in \eqref{eq:prelim_sc_simple} is formally reminiscent of a $5\times5$ depthwise convolution, it is not well matched to a conventional convolution dataflow. 
The effective operands are regenerated from the phase map at every sampling step rather than stored as stationary weights, and the final output still requires a center-dependent multiply--subtract combination that is not naturally supported by a standard line-buffer pipeline. 
These observations motivate the SA-Kura architecture in Section~\ref{sec:architecture}: a systolic pipeline that combines register-level neighbor reuse, in-dataflow center capture, and a two-dimensional offset-sweep schedule.

\section{SA-Kura Architecture and Systolic Dataflow}
\label{sec:architecture}
\subsection{Architecture Overview}
\label{sec:architecture_overview}

SA-Kura is organized as a dedicated drift coprocessor that accelerates the two-stage computation derived in Section~\ref{sec:prelim_optimization}: neighbor accumulation independent of the center phase, followed by a center-dependent multiply--subtract combination.
The score network is executed by a host processor or by a separate accelerator, while SA-Kura computes the drift-side contribution from the same current phase map.
Because both score and drift are functions of the same current state, this organization allows drift evaluation to proceed in parallel with score evaluation and prevents the drift kernel from remaining a serialized residual bottleneck.

As shown in \refFigure{fig:sa_kura_overview}, SA-Kura consists of Local Memory, Controller, Streamer, Column Buffer, Row Buffer, an $N_h \times N_w$ PE array, Ref Term Unit, Theta Update Unit, and a Random Number Generator.
The input image is partitioned into spatial tiles of size $N_h \times N_w$, each of which is directly mapped onto the PE array.
Accordingly, one array invocation processes $N_hN_w$ center pixels in parallel.

For each tile, the Streamer reads phase values $\theta$ from Local Memory, converts them into $\sin\theta$ and $\cos\theta$, and injects the resulting streams into the array through the Column Buffer and Row Buffer.
Each PE is assigned to one center pixel of the current tile.
During the neighborhood sweep, it accumulates transformed neighbor components to build $S_i$ and $C_i$ in \eqref{eq:prelim_sc_simple}.
When the relative offset reaches $(0,0)$, the center components $\sin\theta_i$ and $\cos\theta_i$ are captured locally rather than accumulated.
After the sweep, the PE forms the unscaled neighborhood core $\tilde{u}_i^{\mathrm{nbr}}$ in \eqref{eq:prelim_nbr_core}, and Theta Update Unit applies the global factor $K(t)/|\mathcal{N}_i|$ to realize the neighborhood term in \eqref{eq:prelim_center_combine}.
This separation is intentional: the spatial reduction is performed inside the PE array, whereas the time-dependent scalar scaling is applied once outside the PE datapath.

The remaining deterministic and stochastic components of \eqref{eq:lc_kuramoto} are handled outside the PE array.
Using the same captured center components, the Ref Term Unit computes the reference-attraction term in \eqref{eq:prelim_ref_combine}, and the Random Number Generator provides the sampled stochastic term.
Theta Update Unit combines these with the scaled neighborhood term to produce the drift-side contribution of the discretized update.
In parallel, the score-side term is computed by the host processor or by a separate score-network accelerator.
The Controller then merges the externally computed score term with the drift-side contribution to form $\theta_{\text{new}}$, which is written back to Local Memory\LKW{---provisioned as $N_h$ banks 
of $1024\times32$-bit (4\,kB) SRAM---}after all tiles of the current sampling step have been processed.

\begin{figure}[t]
    \centering
    \includegraphics[width=\columnwidth]{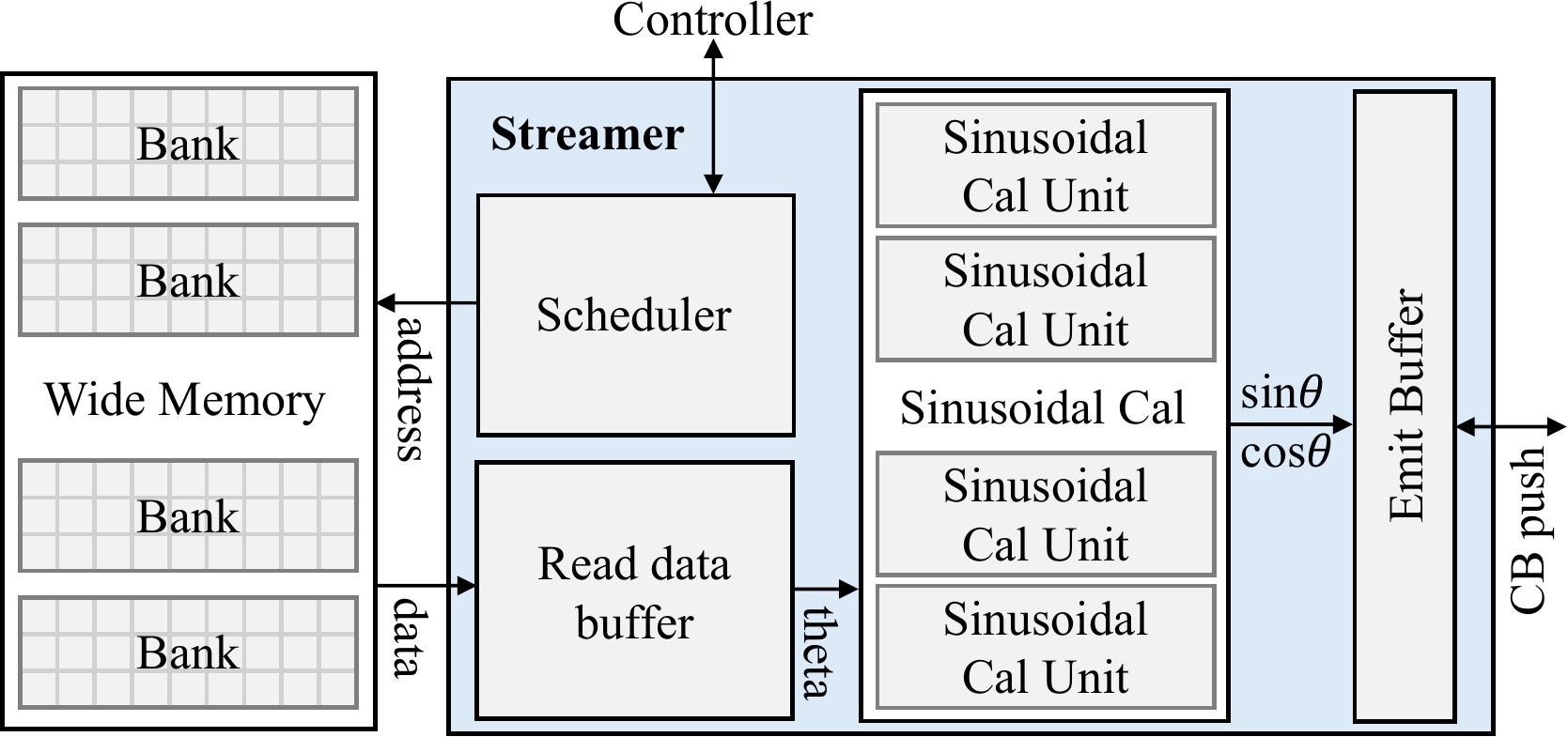}
   \vskip -6pt
\caption{Overview of the Streamer. Phase values are read from wide memory, converted into \boldmath$\sin\theta$ and $\cos\theta$ by parallel sinusoidal units, and buffered before Column Buffer input.}   
\label{fig:streamer}
    \vskip -2pt
\end{figure}

The right panel of \refFigure{fig:sa_kura_overview} shows the PE microarchitecture.
Each PE receives $\sin\theta$ and $\cos\theta$ components from the left and top through \texttt{sin\_L}/\texttt{cos\_L} and \texttt{sin\_U}/\texttt{cos\_U}, and forwards them to the right and downward through \texttt{sin\_R}/\texttt{cos\_R} and \texttt{sin\_D}/\texttt{cos\_D}.
The center components are stored in the Cos Center Buffer (CCB) and Sin Center Buffer (SCB), which hold $\cos\theta_i$ and $\sin\theta_i$, respectively.
Neighbor inputs are accumulated along the ACC path to form $S_i$ and $C_i$.
CRB/SRB align and retain the right-going components during horizontal advancement, while CDB/SDB buffer the downward-propagating components.
Once all neighborhood positions have been visited, the accumulated sums and stored center components are combined through a multiply--subtract datapath to generate $\tilde{u}_i^{\mathrm{nbr}}$.
These values are then forwarded to the array output edge and drained through the split output ports \texttt{drain\_H} and \texttt{drain\_T}.

\subsection{Dataflow-Centric Operation of SA-Kura}
\label{sec:dataflow_operation}


The phase map is stored in Local Memory using a 16-bit Q1.15 representation obtained by normalizing the physical phase range $[-\pi,\pi)$ to $[-1,1)$.
\LKW{
The resulting phase resolution is $\pi\cdot2^{-15}\approx 9.6\times10^{-5}$\,rad, more than two orders of magnitude below the stochastic-term amplitude in the target noise schedule; its 15 fractional bits also exceed the 10-bit mantissa of IEEE\,FP16.
}
As shown in \refFigure{fig:streamer}, the Streamer issues read addresses to a multi-bank wide memory, buffers the returned phase values, and dispatches them to multiple parallel sinusoidal calculation units.
We considered three implementation options for this front-end conversion: CORDIC, direct LUT, and LUT with linear interpolation.
CORDIC offers good accuracy but requires iterative stages and a deeper pipeline, whereas a direct LUT has a short critical path but consumes substantial storage for the target precision.
SA-Kura therefore adopts a quarter-wave LUT over $[0,\pi/2]$ with 4096 stored sine samples and 2-bit fractional linear interpolation.
Combined with quadrant decoding, this design sustains a continuous $\sin\theta/\cos\theta$ stream at low storage cost and removes transcendental units from the PE array.

For a tile of size $N_h \times N_w$, each PE must observe an $M \times M$ neighborhood with $M=5$.
The streamed input therefore includes a halo-extended field of size $(N_h+M-1)\times(N_w+M-1)$.
SA-Kura delivers this field through a two-dimensional offset-sweep schedule over relative offsets $(\Delta x,\Delta y)\in\{-2,-1,0,1,2\}^2$.
The Column Buffer stores $N_w$ columns of transformed data, each of height $N_h+M-1$.
When one column is popped, the upper $M-1$ entries are injected into the left side of the Row Buffer, while the lower $N_h$ entries are injected directly into the left edge of the PE array.
The Row Buffer, placed above the array and sized $(M-1)\times N_w$, holds the extra halo rows required for subsequent downward sweeps.

Before the offset sweep begins, the Row Buffer and PE array are initialized through a prefill phase.
Over $N_w$ cycles, one new transformed column is inserted per cycle from the Column Buffer.
The upper halo entries fill the Row Buffer, while the lower tile-aligned entries fill the PE array, with existing data shifting one position to the right at each cycle.
After these $N_w$ prefill cycles, each PE contains the transformed sample corresponding to the first offset $(\Delta x,\Delta y)=(-2,-2)$ for its assigned center position, and the Row Buffer holds the remaining top-halo rows needed to continue the sweep.

The neighborhood traversal then proceeds as a 2-D offset sweep.
For a fixed horizontal offset $\Delta x$, the Row Buffer shifts downward once per cycle and feeds its top row into the PE array.
Inside the array, the incoming $\sin\theta$ and $\cos\theta$ values propagate systolically from top to bottom, and each PE updates the corresponding accumulations for its assigned center.
After $M$ downward shifts, all vertical offsets $\Delta y=-2,\ldots,+2$ for the current $\Delta x$ have been visited.

To advance to the next horizontal offset, a new transformed column must be inserted from the left while preserving the state required for the next downward sweep.
SA-Kura handles this using CRB/SRB together with the Row Buffer state.
Before each downward sweep, the design snapshots the right-going edge state needed for the subsequent horizontal advancement.
After the sweep, this state is restored, a new leftmost column is injected from the Column Buffer, and the internal contents shift right by one position.
Repeating this horizontal-then-vertical pattern for $\Delta x=-2,\ldots,+2$ causes every PE to visit the full $5\times5$ neighborhood while keeping the center pixel fixed in place and moving only the transformed input field through the buffer/array fabric.

The center components are captured in-dataflow.
Specifically, when the relative offset reaches $(0,0)$, the arriving $\sin\theta_i$ and $\cos\theta_i$ are written into SCB/CCB rather than accumulated.
This does not alter the mathematical result because the self-interaction term is zero.
The remaining $M^2-1$ offsets are accumulated normally into $S_i$ and $C_i$.
After the full sweep is complete, each PE combines its stored center components and accumulated sums to generate $\tilde{u}_i^{\mathrm{nbr}}$.
These values propagate to the PE array output edge and are drained through the split output ports \texttt{drain\_H} and \texttt{drain\_T}.

The post-array datapath completes the drift-side computation.
Theta Update Unit scales each drained value by $K(t)/|\mathcal{N}_i|$ to form the neighborhood term in \eqref{eq:prelim_center_combine}.
In parallel, the Ref Term Unit uses the captured center components to compute the reference-attraction term in \eqref{eq:prelim_ref_combine}.
The scalar coefficients $K_{\mathrm{ref}}(t)\sin\psi_{\mathrm{ref}}$ and $K_{\mathrm{ref}}(t)\cos\psi_{\mathrm{ref}}$ are loaded once per sampling step into configuration registers and are shared across all pixels, so the reference path requires no additional memory access beyond the center components already captured in the PE array.
Theta Update Unit then combines the scaled neighborhood term, the reference term, and the sampled stochastic term from the Random Number Generator to produce the drift-side update.
The score-side term is computed in parallel outside SA-Kura, and the two are merged to form $\theta_{\text{new}}$ before write-back to Local Memory.


Because practical images are larger than the PE array, SA-Kura processes multiple tiles sequentially.
To avoid idle cycles at tile boundaries, the accelerator overlaps the drain of the current tile with the prefill of the next tile.
With this drain--prefill overlap, the steady-state cycle count per tile becomes
\begin{equation}
C_{\text{tile}} = N_w + M^2 + 1,
\label{eq:ctile}
\end{equation}
where the $N_w$ term corresponds to the prefill depth, the $M^2$ term corresponds to the full neighborhood sweep, and the final cycle accounts for center combination and drain.
Since $M$ is fixed by the neighborhood size, the only variable component of $C_{\text{tile}}$ is the array width $N_w$, which directly shapes the throughput--utilization tradeoff explored in the next subsection.

\subsection{Analytical Design-Space Exploration}
\label{sec:dse}

The PE array shape of SA-Kura, parameterized by $(N_h,N_w)$, jointly determines throughput, energy efficiency, and silicon area.
The analytical model in this subsection assumes a common target clock frequency $f_{\text{clk}}$ across configurations, thereby isolating the effect of array shape from configuration-dependent timing closure.
From \eqref{eq:ctile}, the steady-state cycle count per tile is $C_{\text{tile}}=N_w+M^2+1$.
Since one tile produces $N_hN_w$ output pixels, the corresponding pixel throughput is
$\displaystyle
T_{\text{px}} =
\frac{f_{\text{clk}}\,N_hN_w}{N_w + M^2 + 1},
$ 
and the equivalent cycles per pixel are
$\displaystyle
c_{\text{px}} =
\frac{C_{\text{tile}}}{N_hN_w}
=
\frac{1}{N_h}
+
\frac{M^2+1}{N_hN_w}.
$
The latter immediately shows that array height and width affect throughput differently.
In this equation, increasing $N_h$ reduces both terms, whereas increasing $N_w$ reduces only the second term associated with sweep overhead.
Therefore, $(N_h,N_w)$ is not merely a parallelism parameter, but a first-order architectural design variable.


Throughput alone, however, does not determine the best operating point.
Increasing $N_h$ or $N_w$ also enlarges the front-end and buffering structures outside the PE array.
To compare configurations independently of image size, we therefore use the energy per output pixel, $E_{\text{px}}$, as the primary efficiency metric.
Based on the architecture in Section~\ref{sec:architecture_overview}, the total power is modeled as
\begin{equation}
P_{\text{total}} =
p_{hw}N_hN_w
+
p_wN_w
+
p_hN_h
+
p_{\text{fixed}},
\label{eq:psys}
\end{equation}
where $p_{hw}$ captures the per-PE power, $p_w$ the width-scaling overhead, $p_h$ the height-scaling overhead, and $p_{\text{fixed}}$ the array-independent fixed power.
The energy per output pixel is then
$\displaystyle
E_{\text{px}} =
\frac{P_{\text{total}} \cdot C_{\text{tile}}}
     {N_hN_w f_{\text{clk}}}.
$
Substituting \eqref{eq:psys} and \eqref{eq:ctile} into this $E_{\text{px}}$ equation yields
\begin{equation}
E_{\text{px}} =
\frac{N_w+M^2+1}{f_{\text{clk}}}
\left(
p_{hw}
+
\frac{p_w}{N_h}
+
\frac{p_h}{N_w}
+
\frac{p_{\text{fixed}}}{N_hN_w}
\right).
\label{eq:epx_expanded}
\end{equation}
This expression makes the shape dependence explicit.
For a fixed $N_w$, $E_{\text{px}}$ decreases monotonically with $N_h$ because larger array height amortizes width-scaling and fixed overheads over more output pixels.
For a fixed $N_h$, the dependence on $N_w$ is convex: increasing $N_w$ improves parallelism and amortizes the fixed sweep overhead $M^2+1$, but also increases the prefill depth linearly.
Setting $\partial E_{\text{px}}/\partial N_w = 0$ yields the energy-optimal width for a given height:
\begin{equation}
N_w^{*}(N_h)
=
\sqrt{
(M^2+1)\cdot
\frac{p_h + p_{\text{fixed}}/N_h}
     {p_{hw} + p_w/N_h}
}.
\label{eq:nwopt}
\end{equation}
\eqref{eq:nwopt} can be interpreted as the balance point between width-dependent prefill overhead and the parallelism benefit of a wider array.

The silicon area is modeled in the same separable form:
\begin{equation}
A =
a_{hw}N_hN_w
+
a_wN_w
+
a_hN_h
+
a_{\text{fixed}}.
\label{eq:area}
\end{equation}
Here, $a_{hw}$, $a_w$, $a_h$, and $a_{\text{fixed}}$ denote the per-PE, width-scaling, height-scaling, and fixed area coefficients, respectively.
Under an area budget $A_{\text{budget}}$, \eqref{eq:nwopt} and \eqref{eq:area} jointly define the feasible region.
Since $E_{\text{px}}$ decreases monotonically with $N_h$ for any fixed $N_w$, its minimum over $N_w$ also decreases with $N_h$.
Therefore, under the analytical model, the minimum-energy point is attained at the largest feasible $N_h$, together with the corresponding width $N_w^{*}(N_h)$.


The model above intentionally captures the compute-dominated behavior of SA-Kura and excludes transfer overhead.
This isolates the effect of $(N_h,N_w)$ on on-core efficiency while preserving the closed-form insight of \eqref{eq:nwopt}.
Once timing closure and off-core transfer cost are included, the optimum becomes platform-dependent and generally loses this closed-form structure.
We therefore use \refEqn{eq:psys}--\refEqn{eq:area} as the architecture-level design-space model and fit its coefficients using synthesis and RTL results in Section~\ref{sec:eval_synthesis}.

\section{Implementation and Evaluation}
\label{sec:evaluation}

\subsection{Prototype and Setup}
\label{sec:eval_setup}

\textcolor{black}{We implemented SA-Kura as a configurable Verilog RTL accelerator and integrated it into a lightweight SoC platform built around the Rocket RISC-V core~\cite{Rocket}. The SoC integration was performed using RISC-V eXpress (RVX), an EDA tool widely adopted for developing processors on the RISC-V platform~\cite{Han:IoT2021,Park:TCASI24,Lee:IoTJ25,CHOI:AEJ25,Jeon:DAC25,kwak:DATE26}.
The SoC includes a lightweight NoC~\cite{Han:TCAD18}}, system SRAM, a DMA engine, and peripheral interfaces.
The DMA transfers phase-map tiles between system memory and the local memory of SA-Kura over a 128-bit interconnect.
Using this common platform, we instantiated 25 SA-Kura processor configurations with
$(N_h,N_w)\in\{5,10,15,20,25\}\times\{5,10,15,20,25\}$.
For functional validation, the integrated processors were prototyped on the Xilinx VCU118 FPGA board (Virtex UltraScale+)\JJM{~\cite{VCU118}} using Vivado~\cite{Vivado} and operated at 100\,MHz.
On the FPGA prototypes, we verified the locally-coupled Kuramoto drift computation using $96\times96$ single-channel SOCOFing inputs~\cite{SOCOFing:arXiv18}.
Because SA-Kura processes images tile by tile and its local memory is parameterized independently of array shape, the architectural trends reported here are not tied to this image size; larger inputs can be supported by increasing SRAM depth and processing more tiles.

\begin{figure}[t]
    \centering
    \includegraphics[width=\columnwidth]{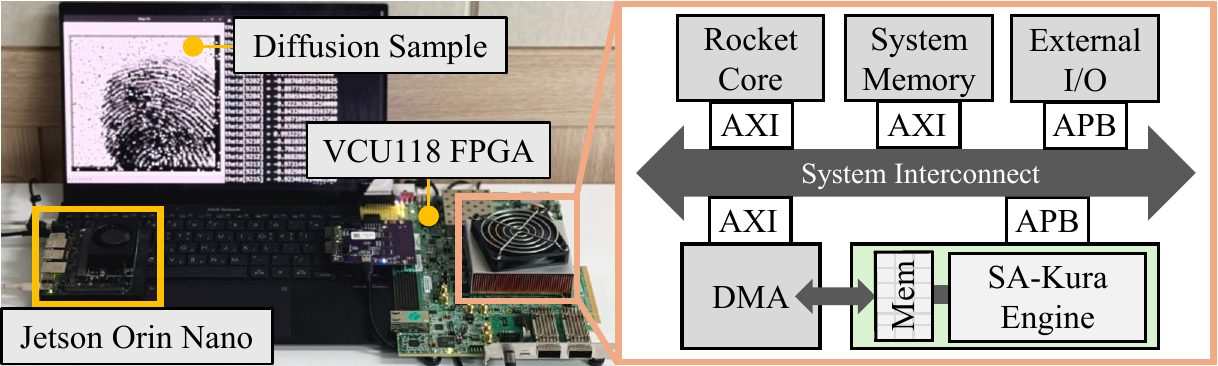}
    \vskip -8pt
\caption{System-level demonstration. Left: heterogeneous setup with a Jetson Orin Nano host and an SA-Kura FPGA prototype on VCU118. Right: prototype architecture.}
\label{fig:demo}
\end{figure}

\begin{table}[t]
\caption{FPGA resource usage and post-synthesis area/power for the H20W5 SA-Kura processor (45\,nm, 100\,MHz).}
\vskip -9pt
\centering
\resizebox{1\columnwidth}{!}{
        \renewcommand{\arraystretch}{1.15}
        \begin{tabular}{l| r r r r | r r}
          \Xhline{1pt}
          \rowcolor[HTML]{D9D9D9}
             Processor IPs & LUTs & FFs & DSPs & BRAM & Area [$\mu$m$^2$] & Power [mW] \\ \hline
             Rocket Core       & 15,228 & 9,755  & 4   & 12 & 584,432   & 10.90 \\
             Peripherals incl.\ DMA & 7,970 & 9,154 & -- & --  & 104,688   & 4.72 \\
             System Interconnect & 9,586 & 11,890 & --  & --  & 127,100   & 11.07 \\
             System Memory       & 174   & 342    & --   & 16 & 823,984   & 3.69 \\ 
            \rowcolor[HTML]{EFEFEF}
             \textbf{SA-Kura}  & \textbf{31,263} & \textbf{28,531} & \textbf{321} & \textbf{20} & \textbf{1,884,657} & \textbf{54.12} \\
             \quad $\llcorner$ Controller          & 539    & 190    & --  & -- & 3,261     & 0.224 \\
             \quad $\llcorner$ Streamer \& Buffers & 5,946  & 6,324  & 1   & -- & 63,379    & 11.12 \\
             \quad $\llcorner$ PE-Array            & 19,420 & 12,520 & 200 & -- & 505,118   & 20.08 \\
             \quad $\llcorner$ Post-array Datapath$^\dagger$ & 5,336 & 9,477 & 120 & -- & 282,919 & 15.46 \\
             \quad $\llcorner$ Local Memory & 22 & 20 & -- & 20 & 1,029,980 & 7.24 \\
            \Xhline{1pt}
            \multicolumn{7}{l}{\footnotesize $^\dagger$Includes Ref Term Unit and Random Number Generator.} \\
        \end{tabular}
    }
\vskip -4pt
\label{tab:fpga_and_ASIC_breakdown}
\end{table}
For quantitative evaluation, all 25 configurations were synthesized using Synopsys Design Compiler~\cite{DesignCompiler} with the Nangate 45\,nm library~\cite{Nangate45} under a common 100\,MHz target.
Dynamic power was estimated in PrimeTime PX~\cite{PrimeTime} using VCD-based switching activity generated from the same $96\times96$ workload.
All experiments use a $5\times5$ neighborhood, 16-bit Q1.15 phase inputs, and 32-bit accumulation; all reported area and power results are obtained from post-synthesis ASIC results, and every configuration satisfies the 100\,MHz timing constraint.
To quantify the benefit of dedicated drift offloading within the intended coprocessor setting, we also evaluate a processor-only baseline in which the Rocket core executes the same fixed-point Kuramoto drift kernel in software, using identical Q1.15 inputs, 32-bit accumulation, and LUT/interpolation-based sinusoidal evaluation. 
This baseline uses the same SoC shell, memory hierarchy, and interconnect as the accelerator-assisted configuration, isolating the effect of SA-Kura itself.

Figure~\ref{fig:demo} shows the heterogeneous demonstration platform, where a Jetson Orin Nano host evaluates the score network while the SA-Kura processor on the VCU118 computes the drift term in parallel.
This host role of Jetson is distinct from the standalone GPU baseline in Section~\ref{sec:eval_gpu}, where the same platform is used only to execute the drift kernel.
Table~\ref{tab:fpga_and_ASIC_breakdown} reports the FPGA resource usage and post-synthesis ASIC breakdown for the representative H20W5 configuration. Here, H$h$W$w$ denotes a configuration with $N_h=h$ PE rows and $N_w=w$ PE columns.
The PE array dominates DSP usage and active power, followed by the Streamer and buffers, while the surrounding SoC shell remains relatively lightweight.

\subsection{Architecture-Level Model Validation}
\label{sec:eval_synthesis}

\begin{figure}[t]
    \centering
    \includegraphics[width=\columnwidth]{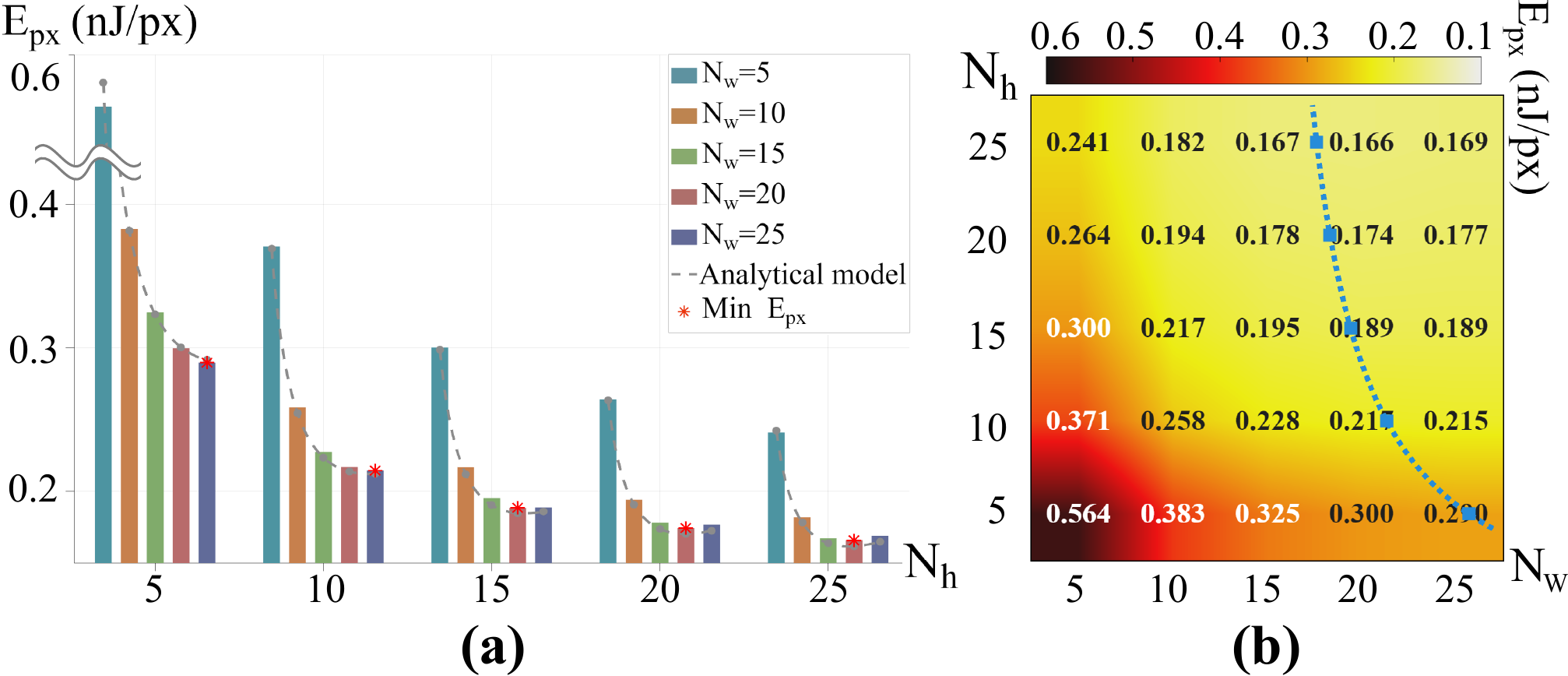}
    \vskip -6pt
\caption{Compute-only \boldmath$E_{\text{px}}$ across 25 configurations. (a) Measured values grouped by $N_h$ with the analytical model. (b) Heatmap with the optimal-width curve $N_w^{*}(N_h)$.}
    \label{fig:st}
\end{figure}

We synthesized and analyzed all 25 array configurations and used the resulting area and power numbers to validate the analytical models derived in Section~\ref{sec:dse}.
Fitting the synthesized results to \refEqn{eq:psys} and \refEqn{eq:area} yields $R^2>0.999$ for both models, confirming that the proposed bilinear decomposition accurately captures the dominant scaling trends of SA-Kura.

Figure~\ref{fig:st} (a) compares the measured compute-only $E_{\text{px}}$ values with the analytical prediction from \refEqn{eq:epx_expanded}, grouped by $N_h$.
Figure~\ref{fig:st} (b) shows the corresponding compute-only $E_{\text{px}}$ heatmap together with the optimal-width curve $N_w^{*}(N_h)$ predicted by \refEqn{eq:nwopt}.
The measured data closely follow the predicted trend: compute-only $E_{\text{px}}$ decreases monotonically with increasing $N_h$, while the low-energy valley is well tracked by the predicted optimal-width curve.

The closed-form cycle model in \refEqn{eq:ctile} is also verified directly against RTL simulation.
For all 25 evaluated configurations, the measured tile-level cycle count exactly matches the prediction of \refEqn{eq:ctile}, confirming that the drain--prefill-overlapped offset-sweep schedule is fully captured by the analytical timing model.

The open markers in \refFigure{fig:pareto} plot the compute-only area--$E_{\text{px}}$ tradeoff.
Under the compute-only metric, larger arrays improve intrinsic energy efficiency because the fixed sweep overhead is amortized over more output pixels.
However, the best explored configuration still depends on the area budget: under area budgets of 3, 4, 5 and 6\,mm$^2$, the best explored configurations are H10W10 (258.5\,pJ/px), H15W15 (195.4\,pJ/px), H20W15 (178.1\,pJ/px), and H25W20 (166.0\,pJ/px), respectively.
Importantly, the compute-only metric is not intended to determine the final deployed operating point by itself; rather, it isolates the architectural effect of array shape before transfer overhead is introduced.

\subsection{System-Level Comparison}
\label{sec:eval_gpu}

While the analytical model isolates intrinsic accelerator efficiency, a deployed processor subsystem also incurs software-side data marshaling and DMA transfer overhead. 
We therefore define system-level $E_{\text{px}}$ as the total energy of drift execution including these costs, while still excluding score-network inference. 
Accordingly, the system-level optimum does not necessarily coincide with the compute-only frontier.

\begin{figure}[t]
    \centering
    \includegraphics[width=\columnwidth]{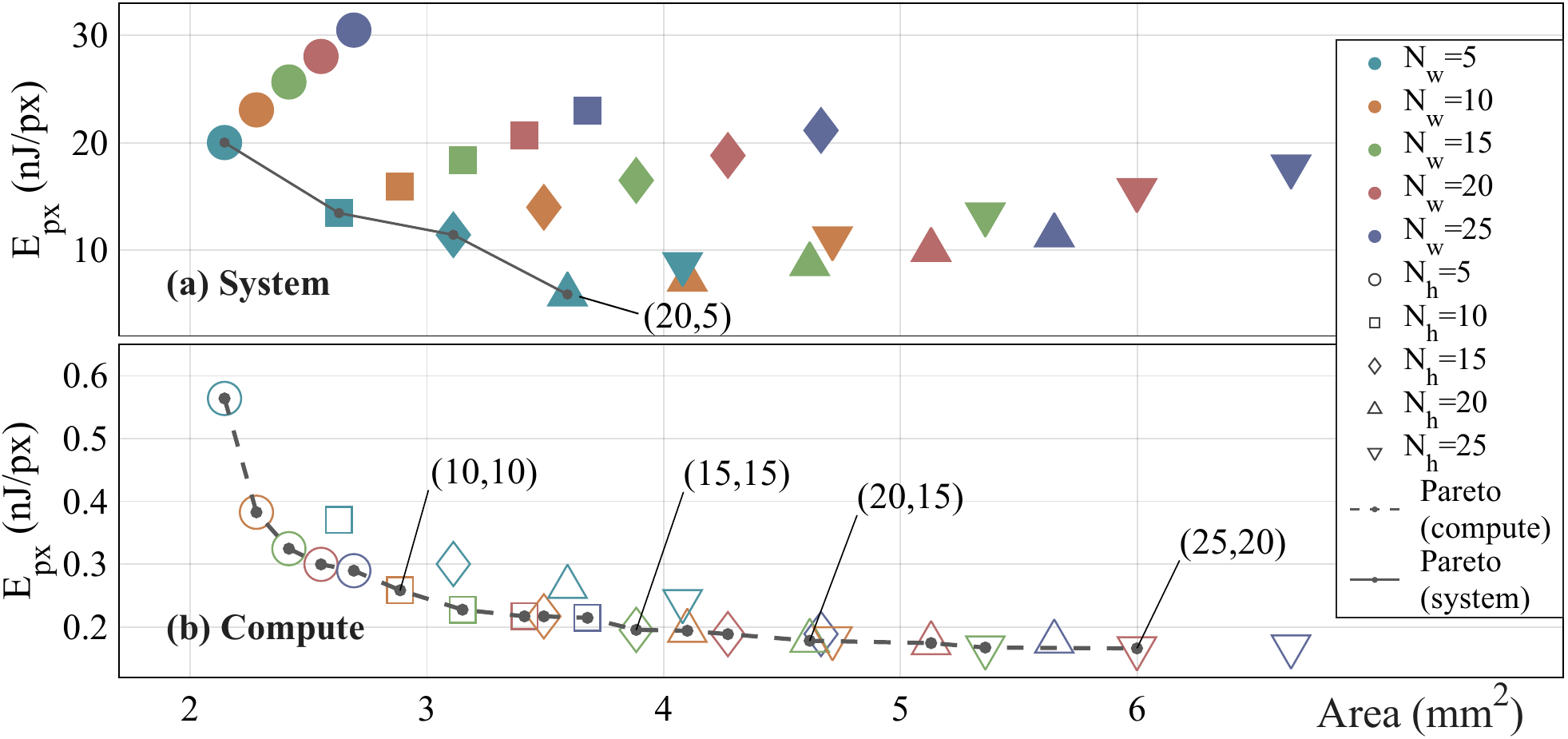}
        \vskip -10pt
\caption{Area--\boldmath$E_{\text{px}}$ tradeoff for 25 SA-Kura configurations.}
    \label{fig:pareto}
\end{figure}

This shift is visible in \refFigure{fig:pareto}, where filled markers denote system-level $E_{\text{px}}$. Large arrays lose part of their advantage because reduced execution time is offset by higher concurrent power during transfer-dominated phases. Among the 25 configurations, the minimum measured system-level energy is achieved by H20W5 at 5.88\,nJ/px. 
\LKW{The system-optimal shape differs markedly from the compute-only frontier.
H20W5 achieves the lowest per-transaction DMA cost because its $N_h{=}20$ banks align exactly to the 128-bit interconnect, minimizing system-level $E_{\text{px}}$ despite requiring more DMA transactions than the largest configurations.}

To quantify the benefit of dedicated offloading, we compared H20W5 against a processor-only baseline where the Rocket core executes the fixed-point Kuramoto drift in software. Because both designs use the same SoC shell, memory system, and fixed-point formulation, this comparison isolates the effect of SA-Kura. The Rocket-only baseline requires 123.8\,ms at 30.38\,mW, whereas H20W5 completes the same task in 641.3\,$\mu$s at 84.5\,mW. Thus, SA-Kura reduces latency by approximately $192.99\times$ and energy by approximately $69.39\times$, despite only $2.78\times$ higher instantaneous power.

For external comparison, we measured the drift computation on a Jetson Orin Nano GPU using a standalone CUDA implementation. Unlike Figure~\ref{fig:demo}, Jetson is used here solely as a baseline for the drift kernel. After idle-power calibration and 1000-run averaging, the measured energy is 270.57\,nJ/px. Relative to this GPU baseline, the system-optimal H20W5 configuration achieves approximately $46.02\times$ lower energy per pixel and $6.57\times$ lower latency per step.
This large gap is structural rather than purely arithmetic: SA-Kura uses a dedicated systolic datapath with register-level reuse matched to the stencil structure, whereas the same workload is poorly matched to general-purpose GPUs. In fact, the measured utilization on Jetson is only 25.1\%, indicating that nonlinear phase interactions remain inefficient on matrix-oriented engines. This supports our central claim that kernels with modest arithmetic counts can become dominant bottlenecks when their patterns are mismatched to the underlying hardware.

\section{Conclusion}


This paper presented SA-Kura, a digital systolic-array accelerator for locally-coupled Kuramoto drift in diffusion sampling. By reformulating the center-dependent sinusoidal stencil into center-phase-independent neighbor accumulation followed by a final center-dependent multiply--subtract combination, SA-Kura eliminates in-PE transcendental units and enables an efficient offset-sweep systolic dataflow with register-level neighbor reuse. Across 25 synthesized configurations in 45\,nm CMOS, the closed-form cycle model matches RTL simulation exactly, and the bilinear power and area models achieve $R^2>0.999$. 
Under system-level evaluation including transfer overhead, the H20W5 configuration achieves the best energy efficiency at 5.88\,nJ/px. Relative to software execution of the same fixed-point drift kernel on the Rocket processor, it reduces latency and energy by approximately $193\times$ and $69.4\times$, respectively. Relative to a Jetson Orin Nano CUDA baseline, it is approximately $6.6\times$ faster and achieves approximately $46\times$ lower energy per pixel.


\bibliographystyle{unsrt}
\bibliography{reference}
\end{document}